\documentclass[12pt]{article}
\usepackage{epsfig}

\newcommand{\beqn}{\begin{eqnarray}}
\newcommand{\eeqn}{\end{eqnarray}}
\newcommand{\vp}{\varphi}
\newcommand{\vpcl}{\varphi_{\mathrm{cl}}}
\newcommand{\vd}{\vec \diff}
\newcommand{\cZ}{{\cal Z}}
\newcommand{\cC}{{\cal C}}
\newcommand{\cL}{{\cal L}}
\newcommand{\dd}{{\mathrm d}}
\newcommand{\dD}{{\mathrm D}}
\newcommand{\diff}{\partial}
\newcommand{\eq}[1]{(\ref{#1})}

\makeatletter
\def\sileqq{\mathrel{\mathpalette\gs@align<}}
\def\sigeqq{\mathrel{\mathpalette\gs@align>}}
\def\gs@align#1#2{\lower.6ex\vbox{\baselineskip\z@skip\lineskip\z@
    \ialign{$\m@th#1\hfil##\hfil$\crcr#2\crcr\sim\crcr}}}
\makeatother

\date{}
\begin{document}
\title{
Confinement in three dimensional magnetic monopole--dipole gas
\vskip-35mm
\rightline{\small ITEP-TH-17/00}
\vskip 30mm
}
\author{M.N.~Chernodub
\\
{\small\it Institute of Theoretical and Experimental Physics,}\\
{\small\it B.Cheremushkinskaya 25, Moscow, 117259, Russia}\\
\vspace{2\baselineskip}
}

\maketitle
\thispagestyle{empty}

\begin{abstract}
Confinement of electrically charged test particles in the dilute
plasma of mo\-no\-po\-les and pointlike magnetic dipoles is studied.
We calculate the tension of the string emerging between the
infinitely separated test particles. The string tension is an
increasing function of the dipole density provided other parameters
of the plasma are fixed. The relevance of our results to confining
gauge theories is discussed.
\end{abstract}

\vskip 1.5 cm

\section{Introduction}

We study the confining properties the plasma of the Abelian magnetic
monopoles with a fraction of the magnetic dipoles in three
dimensional Euclidean space--time. This kind of the plasma may appear
in gauge theories such as the Georgi--Glashow model which possesses
the topologically stable classical solution called the
't~Hooft--Polyakov monopole~\cite{tHPo}. The dipoles are realized as
the monopole--anti-monopole bound states. Since the long range gauge
fields of the monopole are associated with an unbroken Abelian
subgroup the long range properties of the monopoles and as a
consequence, of the dipoles, are Abelian.

The confining properties of the dilute Abelian monopole--antimonopole
plasma are well known~\cite{Polyakov}. The test particles with
opposite electric charges are confined due to formation of a
stringlike object between the charge and the anticharge. The string
has a finite thickness of the order of the inverse Debye mass and a
finite energy per the string length ("string tension"). Thus the
potential between the test particles is linear at large
distances~\cite{Polyakov}.

At a sufficiently high temperature the plasma of the Abelian
monopoles undergoes the Be\-re\-zin\-sky--Kos\-ter\-litz--Thou\-less
phase transition~\cite{B,KT,Parga}. The vacuum in the high
temperature phase is filled with the neutral monopole--anti-monopole
bound states~\cite{KT,FrSp80,AgasianZarembo} obeying non--zero
magnetic dipole moments\footnote{Note that the monopole binding is
qualitatively similar to the formation of the instanton molecules in
the high temperature phase of QCD suggested to be responsible for the
chiral phase transition~\cite{Shuryak}.}. The long range fields of
the magnetic dipoles are weak and they can not induce a
non--zero string tension between the electric charges separated by
large distances.  Thus the finite temperature phase transition
separates the confining and deconfining phases of the model.

Since the formation of the magnetic dipole states is supported by
the increase of the temperature we expect to have a mixed plasma of
the Abelian monopoles and dipoles at the confining side of the phase
transition. In this paper we investigate the behavior of the electric
charges in the mixed plasma. Below we show that the magnetic dipoles
affect the string tension even despite of the inability of the pure
magnetic dipole gas to confine electric charges. For simplicity the
dipoles are considered to be pointlike.

Note that in the finite temperature Georgi--Glashow model the charged 
bosons are supposed~\cite{Alex} to play an important role. In this 
paper we are interested in the pure monopole--dipole gas without any 
charged matter fields. The structure of the paper is as follows. The 
path integral formulation as well as simplest properties of the 
monopole and dipole gas are discussed in Section~\ref{section:path}. 
The interaction of the electric charges are studied both numerically 
and analytically in Section~\ref{section:charges}. Our conclusions 
are summarized in the last Section.

\section{Gas of monopoles and magnetic dipoles}
\label{section:path}

The partition function of the monopole--dipole gas has the 
following form:
\beqn
\label{eq:Z}
& \cZ & = \sum\limits^\infty_{M=0} \frac{\zeta^M_m}{M!}
\Biggl[\prod\limits^M_{a=1} \Bigl(\sum\limits_{q_a}
\int \dd^3 x_a \Bigr)\Biggr]
\sum\limits^\infty_{N=0} \frac{\zeta^N_d}{N!}
\Biggl[\prod\limits^N_{\alpha=1} \Bigl(
\int\limits_{{\vec n}^2_\alpha = 1} \dd n_\alpha \,
\int\limits^\infty_0 \dd r_\alpha \, F(r_\alpha) \,
\int \dd^3 x_\alpha \Bigr)\Biggr] \\
& & \exp\Bigl\{ - \frac{1}{2} \int \dd^3 x \int \dd^3 y
\Bigl[\rho(x) + ({\vec \rho}^{(\mu)}(x), \vd_x) \Bigr] \,
\Bigl[\rho(y) + ({\vec \rho}^{(\mu)}(y), \vd_y) \Bigr] \,
D_{\mathrm{reg}}(x-y)\Bigr\}\,, \nonumber
\eeqn
where $\zeta_m$ ($\zeta_d$) is the fugacity of the monopole (dipole)
component of the gas, and
\beqn
\rho(x) = g_m \sum\limits_a q_a \, \delta^{(3)} (x - x_a)\,,
\quad
{\vec \rho}^{(\mu)}(x) = \sum\limits_\alpha {\vec \mu}_\alpha \,
\delta^{(3)} (x - x_\alpha)\,,
\eeqn
are the densities of the magnetic charges and magnetic dipole
moments, respectively. We use the latin, $a$ (greek, $\alpha$)
subscripts to denote the monopole (dipole) parameters. The magnetic
charge of the $a^{\mathrm{th}}$ individual monopole in units of the
fundamental monopole charge, $g_m = 4 \pi \slash g$, is referred to
as $q_a$. In the dilute monopole gas the (anti)monopoles have unit
magnetic (anti)charges, $|q_a| = 1$.

The magnetic moment of the $\alpha^{\mathrm{th}}$ dipole is ${\vec
\mu}_\alpha = g_m\, {\vec n}_\alpha \, r_\alpha$ (no sum is
implemented), where $r_\alpha$ is the "dipole size" and ${\vec
n}_\alpha$ is the direction of the dipole magnetic moment. For
simplicity we consider a dilute gas of the point--like dipoles
characterized only by a (fluctuating in general case) dipole moment
and the space--time position. At the end of the paper we discuss the
applicability of the results obtained in the pointlike
dipole approximation to the real physical systems.

In eq.\eq{eq:Z} the integration over the dipole moments $\int \dd^3
\mu_\alpha$ is represented as the integration over direction ${\vec
n}_\alpha$ of the dipole moment and over the dipole size $r_\alpha$,
weighted with a normalized distribution function $F(r)$,
$\int^\infty_0 F(r) \, \dd r = 1$.  The function $D(x)= {4 \pi
|x|}^{-1}$ in eq.\eq{eq:Z} is the inverse Laplacian, and the
subscript "reg" indicates that the monopole and dipole
self--interaction terms are subtracted.

The second line of eq.\eq{eq:Z} can be rewritten as
follows\footnote{Here and below we omit inessential constant
factors in front of the path integrals.}:
\beqn
\int \dD \vp \, \exp\Bigl\{ - \frac{g^2}{32 \pi^2} \int \dd^3 x \,
{(\vd \vp)}^2
+ i \sum\limits_a q_a \vp(x_a)
+ i \sum\limits_\alpha (\vec r_\alpha,\vd)
\vp(x_\alpha)
\Bigr\}\,.
\label{eq:inter}
\eeqn
Substituting this equation into eq.\eq{eq:Z}, performing summation
over the monopole charges $q_a$ and integrating over the dipole
moment directions $\vec n_\alpha$, and the monopole (dipole),
$x_{a(\alpha)}$ positions, we get the following partition function
for the gas of the monopoles and dipoles:
\beqn
\label{Zfin}
\cZ & = & \int \dD \vp \, \exp\Bigl\{ - \int \dd^3 x \,
\cL(x) \Bigr\}\,, \\
\cL & = & \frac{g^2}{32 \pi^2} {(\vd \vp)}^2
- 2 \zeta_m \, \cos\vp
- 4 \pi \zeta_d \, \int\limits^{\infty}_0 \dd r \, F(r) \,
\frac{\sin (r\, |\vd \vp|)}{r\, |\vd \vp|}\,,
\label{Lfin}
\eeqn
where $|\vec a| = \sqrt{{\vec a}^2}$.

We study the magnetic monopole--dipole plasma in the weak coupling
regime. The density of the monopoles and antimonopoles, $\rho_m$,
and the density of the dipoles, $\rho_d$, can be calculated as the
following expectation values taken in the statistical sum \eq{eq:Z}:
$\rho_m = <\! M \!>$ and $\rho_d = <\! N \!>$. In the weak coupling
limit the leading order contributions to the (anti)monopole and
dipole densities are proportional to the corresponding
fugacities~\cite{Polyakov,Ch00}:
\beqn
\rho_m = 2 \zeta_m\,,\quad \rho_d = 4 \pi \zeta_d\,.
\label{densities}
\eeqn

The correlations of the fields $\vp$ in model \eq{Lfin} are
short ranged indicating that the photon
gets a non-zero mass in the presence of the monopole
plasma~\cite{Polyakov}. Indeed the expansion of lagrangian \eq{Lfin}
in small fluctuations of the field $\vp$ gives:
\beqn
\cL = \frac{g^2}{32 \pi^2} {\Biggl(1 + \frac{4 \pi}{3} \zeta_d \,
{\overline{ \mu^2}}\Biggr)} {(\vd \vp)}^2
+ \zeta_m \, \vp^2 + O(\vp^4)\,,
\label{Lfin2}
\eeqn
where
\beqn
{\overline{\mu^2}} = g^2_m \, {\overline{ r^2}} =
\frac{16 \pi^2}{g^2} \int\limits^\infty_0 \dd r \, F(r) \, r^2\,,
\eeqn
is the mean quadratic magnetic moment of a dipole. Thus the
two--point
correlation function $<\vp(0) \,\vp(x)>$ is exponentially
suppressed at large distances as $e^{- m_D\, |x|}$, where $m_D$ is
the Debye mass in the monopole--dipole plasma:
\beqn
m^2_D = \epsilon^{-1} \, M^2_D \,,\quad
M^2_D = \frac{32 \pi^2 \zeta_m}{g^2}\,.
\label{mD}
\eeqn
Here $M_D$ is the Debye mass in the pure magnetic monopole plasma and
$\epsilon$ is the dielectric constant (permittivity) of the magnetic
plasma,
\beqn
\epsilon = 1 + \frac{4 \pi}{3} \zeta_d \,{\overline{ \mu^2}}\,.
\label{epsilon}
\eeqn
The presence of the dipole states in the plasma leads to charge
renormalisation, $g^2 \to \epsilon \, g^2$ according to
eqs.(\ref{Lfin2},\ref{mD}). In the next Section we show that the
dipoles affect also the confining properties of the monopole plasma.

\section{Electric charges in magnetic monopole--dipole gas}
\label{section:charges}

Let us consider two static infinitely heavy electrically charged test
particles in the pure magnetic monopole plasma. The monopoles and
anti-monopoles form a double layer at the sides of the minimal
surface, spanned on the particles' trajectories~\cite{Polyakov}: at
one side of the surface the magnetic density is positive while at
another side the density is negative. The stringlike structure
between the charges has a finite thickness of the order of the
inverse Debye mass, $M^{-1}_D$, and a non-zero energy density per
unit area of the surface. Thus, the electric charges are confined in
the magnetic monopole plasma. At large separations $R$ between the
test particles the confining potential is linear \cite{Polyakov}
while at $R M_D \ll 1$ the potential becomes of the form $const. \,
R^\alpha$, where $\alpha \approx 0.6$, Ref.~\cite{ChGuPoZa:GG}.

The behavior of the interparticle potential in the pure plasma of
pointlike dipoles is different. The potential is linear at small
interparticle separations, $V(R) = {\mathrm{const.}}\, \rho_d \,
{\bar r}\, R$, due to interaction of the dipole clouds gathered near
the electric sources \cite{Ch00}. At large distances $R$ the
potential is of the Coulomb type with the renormalized coupling $g$,
$g^2 \to \epsilon g^2$, where $\epsilon$ is the dielectric constant
given in eq.\eq{epsilon}.

Thus, the pure gas of pointlike dipoles gives rise to the nontrivial
modification of the interparticle potential only at small separations
$R$ since the fields of an individual dipole is decreasing faster
than the monopole field and we could expect only a short
range modification of the interparticle potential. However, in the
mixed magnetic dipole--monopole plasma the role of the dipoles at
large distances becomes nontrivial. Physically, the monopole fraction
of the plasma leads to the formation of the stringlike structure
between electrically charged particles while the dipole fraction of
the plasma interacts with the electric field inside the string. This
is the way how the dipoles modify the linear term of the
interparticle potential at large distances. Below we study the
influence of the dipole fraction of the gas on the long--range
potential between electrically charged particles.

Let us consider the contribution of the monopole--dipole gas into the
quantum average of the Wilson loop operator, $W_\cC$, where contour
$\cC$ is the trajectory of the infinitely heavy test particle
carrying the fundamental electric charge, $g \slash 2$:
\beqn
W_\cC = \exp\Bigl\{i \, \frac{g}{2}\, \int \dd^3 x \, \Bigl(\rho(x) +
({\vec \rho}^{(\mu)} (x), \vd_x)\Bigr)
\eta^\cC(x) \slash (2 \pi)\Bigr\}\,.
\label{WL}
\eeqn
The function $\eta_\cC$ is defined as follows:
\beqn
\label{eta:gen}
\eta^\cC(x) = \pi \int\limits_{\Sigma_\cC} \dd^2 \sigma_{ij}(y) \,
\varepsilon_{ijk} \, \partial_k D(x-y)\,,
\eeqn
where the integration is taken over an arbitrary
surface\footnote{Note that the Wilson loop does not depend on the
shape of this surface.} $\Sigma_\cC$ spanned on the contour $\cC$.

Substituting eq.\eq{WL} into eq.\eq{eq:Z} and
performing the transformations presented in
Section~\ref{section:path} we derive the following representation for
the contribution of the monopoles and dipoles to the Wilson loop:
\beqn
{<W_\cC>}_{\mathrm{gas}} & = & \frac{1}{\cZ} \int \dD \vp \,
\exp\Bigl\{ - \int \dd^3 x \, \Bigl[\frac{g^2}{32\, \,\pi^2}
{\Bigl(\vec \partial \vp - \vec \partial \eta^\cC(x)\Bigr)}^2
+ 2 \zeta_m \Bigl(1 - \cos \varphi \Bigr) \nonumber\\
& & + 4 \pi \zeta_d\,
\int\limits^\infty_0 \dd r \, F(r) \,
\Bigl( 1 - \frac{\sin(r \,|\vec \partial \vp|)}{
r \,|\vec \partial \vp|}\Bigr)\Bigr]\Bigr\}\,,
\label{WL:quantum}
\eeqn
where the expression in the square brackets has been normalized to zero
at $\vp=0$ and we have shifted the field $\vp \to \vp - \eta^\cC$.

In the weak coupling regime the densities of the monopoles are small
and the Wilson loop in the leading order is given by the classical
contribution to eq.\eq{WL:quantum}. The corresponding
classical equation of motion is:
\beqn
& & {\vd}^2 \vp - M^2_D \, \sin \vp
- \frac{128 \pi^3 \zeta_d}{g^2} \int\limits^\infty_0 \dd r \, F(r)
{\vd} \Biggl[\frac{{\vd} \vp}{{(\vd \vp)^2}}
\Biggl( \cos(r |\vd \vp|) - \frac{\sin (r |\vd \vp|)}{r |\vd \vp|}
\Biggr)\Biggr] \nonumber\\
& = & \pi \int\limits_{\Sigma_\cC} \dd^2 \sigma_{ij}(y) \,
\varepsilon_{ijk} \, \partial_k \delta^{(3)}(x-y)\,.
\label{class1}
\eeqn

Below we consider the infinitely separated static charge and
anticharge located at the points $(y,z) = (\pm\infty,0)$. The
$x$--coordinate is considered as a time coordinate. The string
$\Sigma_\cC$ in eq.\eq{eta:gen} is chosen to be flat and is defined
by the equation $z=0$. Thus, the classical solution of
eq.\eq{class1} does not depend on $x$ and $y$ coordinates, $\vpcl
= \vpcl(z)$. For the sake of simplicity we consider the
case of the non--fluctuating dipole moments, $G(r) = \delta(r - r_0)$
(below we use $r$ instead of $r_0$).

The classical contribution to the Wilson loop
\eq{WL:quantum} gives the area low, ${<W_\cC>}_{\mathrm{gas}} =
{\mathrm {const.}}\, \exp\{- Area\, \sigma_{\mathrm{cl}}\}$, where the
classical string tension is
\beqn
\sigma_{\mathrm{cl}} = \frac{g\, \sqrt{\rho_m}}{4 \pi}
\int\limits^{+\infty}_{-\infty} \dd \xi \, \Biggl[
\frac{1}{2} {\Bigl(\diff_\xi\vpcl
- 2 \pi \delta(\xi)\Bigr)}^2 + 1 - \cos \vpcl
+ s \, \Biggl( 1 - \frac{\sin(h \,|\diff_\xi \vpcl|)}{
h \,|\diff_\xi \vpcl|}\Biggr)\Biggr]\,,
\label{string}
\eeqn
and the $\vpcl$ is the solution of the rescaled
classical equation of motion~\eq{class1}:
\beqn
& & \diff^2_\xi \vpcl - \sin \vpcl
- s\, \diff_\xi \Biggl[\frac{1}{|\diff_\xi \vpcl|}
\Biggl(\cos(h |\diff_\xi \vpcl|)
- \frac{\sin(h |\diff_\xi \vpcl|)}{h |\diff_\xi \vpcl|}
\Biggr)\Biggr] = 2 \pi \diff_\xi \delta (\xi)\,.
\label{class2}
\eeqn
The dipole size in units of the Debye mass is denoted as
\beqn
h = r \, M_D\,,
\label{size:h}
\eeqn
$\xi = z \, M_D$ is the rescaled $z$--coordinate and
\beqn
s = \frac{\rho_d}{\rho_m} \equiv \frac{2\, \pi\, \zeta_d}{\zeta_m}\,,
\label{gamma}
\eeqn
is the ratio of the dipole density to the monopole density ("dipole
fraction"), eq.\eq{densities}.

In the absence of the dipoles, $s=0$, the
solution of eq.\eq{class2} and the corresponding string tension
are~\cite{Polyakov}:
\beqn
\vpcl^{(0)}(\xi) = 4 \,
{\mathrm{sign}}(\xi)\, \arctan e^{-|\xi|}\,,
\quad \sigma^{(0)}_{\mathrm{cl}} = \frac{2 g\, \sqrt{\rho_m}}{\pi}\,,
\label{tension0}
\eeqn
respectively. In the presence of the dipoles the string tension is
modified:  \beqn \sigma_{\mathrm{cl}} = \sigma^{(0)}_{\mathrm{cl}} \,
H(s,h)\,.
\label{H}
\eeqn
We study the function $H$ below.

In the limit of the small dipole size, $h \ll 1$, eq.\eq{class2} can
be expanded in powers of $h$ up to the second order and the classical
string tension can be found analytically. The correction factor $H$
is the square root of the dielectric constant \eq{epsilon},
\beqn
H(s,h) = {\Bigl[1+\frac{s}{3} \, \Bigl(
h^2 + O(h^4)\Bigr)\Bigr]}^{\frac{1}{2}}\,, \quad h \ll 1\,.
\label{H0}
\eeqn
Obviously, this correction is due to the renormalisation of the
coupling constant $g^2 \to \epsilon\, g^2$.

\begin{figure}[!htb]
\vspace{5mm}
 \begin{minipage}{15.0cm}
 \begin{center}
  \epsfig{file=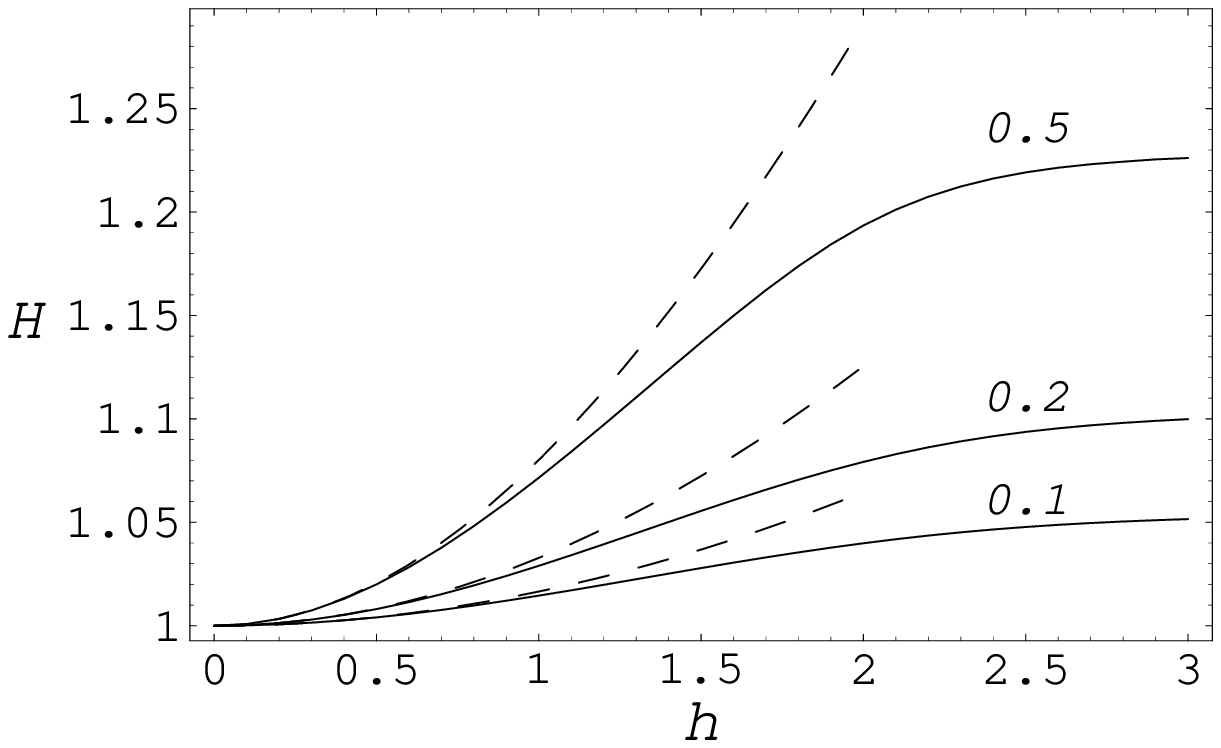,width=7.2cm,height=6.05cm} \hspace{3mm}
  \epsfig{file=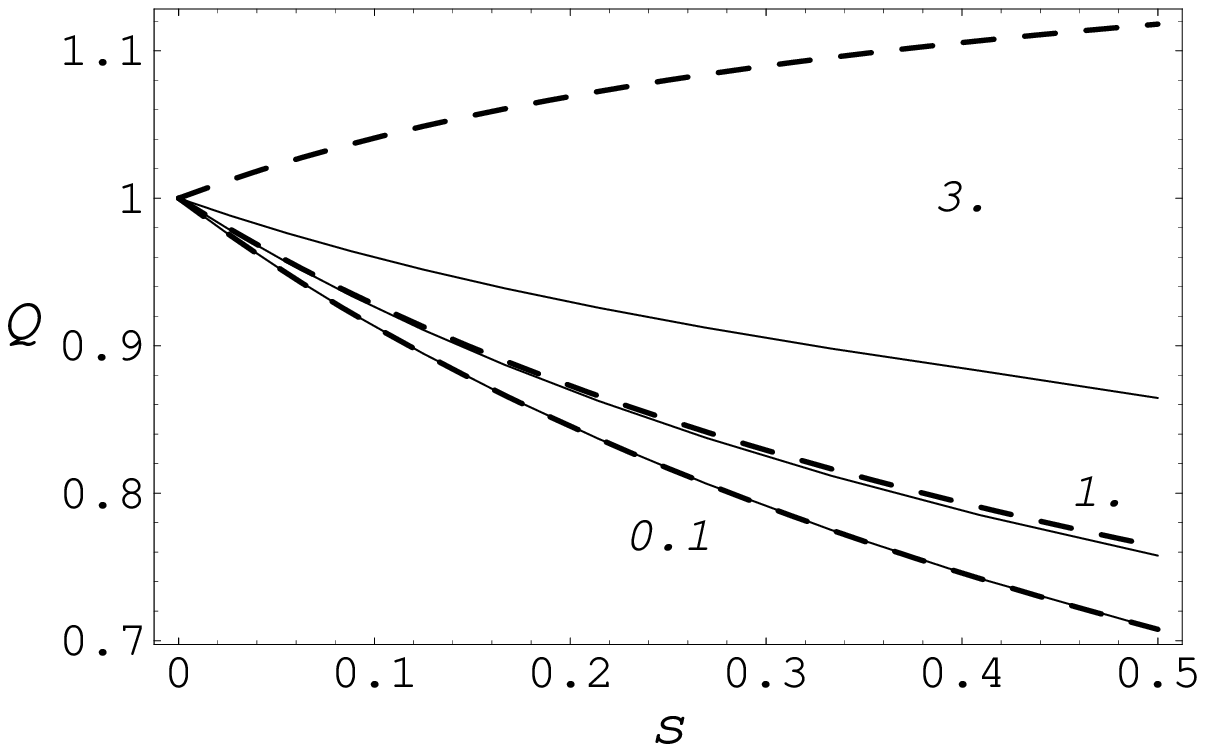,width=7.2cm,height=6.0cm}\\
$(a)$ \hspace{7.0cm} $(b)$\\
 \end{center}
 \end{minipage}
\caption{(a) The correction factor $H$, eq.\eq{H},
to the string tension {\it vs.} the
dipole size $h$ for the different dipole fractions \eq{gamma}:
$s=0.1$,~$0.2$,~$0.5$; (b) the correction factor $Q$, eq.\eq{Q},
$vs.$ the dipole fraction $s$ for the dipole sizes
$h=0.1$, $1$, $3$. The numerical results are shown by the solid lines
and the corresponding leading contributions at small $h$ are shown by
the dashed lines.}
\label{fig1}
\vspace{5mm}
\end{figure}

Unfortunately, eq.\eq{class2} can not be solved analytically for
general values of the dipole fraction $s$ and the dipole size
$h$. Therefore we solve this equation and calculate the correction
factor $H$ numerically. The function $H$ at various (fixed)
dipole fractions $s$ is shown in Figure~\ref{fig1}(a) by solid
lines. The dashed lines indicate the behavior of correction factor
$H$, eq.\eq{H0} valid at the small dipole sizes $h$.

The behavior of function $H$ can be understood as follows. The size
of the flux of the electric field coming from the test
particles is of the order of the inverse Debye mass. The dipoles with
small size $h$ do not feel the flux structure and the
contribution of the dipole fraction of the gas to the string tension
is solely due the polarization of the media of the magnetic dipoles
by the electric field, eq.\eq{H0}. The correction factor $H$ is a
rising function of $h$ since the increase of the dipole magnetic
moment (size) leads to the enhancement of the mean polarization energy
of the dipole in the external field.

At $h \approx 1$ the function $H(h)$ starts to deviate from eq.\eq{H0}
since the dipoles begin to feel the width of the electric flux.
At $h \approx 2$ the rise of the function $H$ slows
down, Figure~\ref{fig1}(a), and at large values of $h$ the correction
factor grows logarithmically, $H \sim s\,  \log h$. To understand
this effect we note that the pure dipole gas can not generate a
finite correlation length~\cite{FrSp80,no-screening} contrary to the
pure monopole gas. Therefore one can expect that the dipole fraction
of the monopole--dipole gas has no essential influence on the Debye
mass\footnote{A numerical study of the profiles of the field
$\vpcl(z)$ shows that this is indeed the
case for all studied values of $h$ and $s$.}. As a result, the
derivative $|\diff_\xi \vpcl|$ is of the order of unity at the
center of the electric flux and the derivative
vanishes out of the flux core
as $e^{- C |\xi|}$, $C \sim 1$. One can easily check that the dipole
contribution to the string tension (the last term in eq.\eq{string})
in the large $h$ limit is equal to $s$ at $|\xi| \sileqq \log h$
and vanishes quickly out of this region. Since in this limit the
other terms in eq.\eq{class2} do not grow with the enlargement of $h$,
the string tension is given by the following formula,
\beqn
\sigma = \frac{s\, g\, \sqrt{\rho_m}}{4 \pi}
\Bigl(\log h + O(1)\Bigr)\,, \quad h \gg 1\,.
\label{sigma:h:large}
\eeqn

Another interesting question is the dependence of the string tension
$\sigma$ on the dipole fraction $s$ at fixed {\it total} number of
the monopoles. The total (anti)monopole number includes both the free
(anti)monopoles and the constituent (anti)monopoles of the magnetic
dipole states. Thus, $\rho_{tot} = \rho_m+2 \rho_d = \rho_m \, (1+ 2
\, s)$, where the dipole fraction $s$ is defined in eq.\eq{gamma}.
If a part of the free monopoles transforms into the dipole states the
string tension acquires positive a contribution due to the increase
of the dipole states (the function $H$ grows) and a
negative contribution due to the decrease of the monopole density
($\sigma^{(0)}_{\mathrm{cl}}$ diminishes). The
correction factor $Q$ due to the monopole binding is defined by the
following equation:
\beqn
\sigma(s,h) = \sigma_{tot} \, Q(s,h) \,,\quad  Q(s,h)
= {(1+2s)}^{-1 \slash 2}\, H(s,h)\,.
\label{Q}
\eeqn
where $\sigma_{tot} = 2 g\, \sqrt{\rho_{tot}} \slash \pi$ is the
string tension in the absence of the dipole states. Note that
$\sigma_{tot}$ remains constant during the binding process.

In Figure~\ref{fig1}(b) we show the function $Q$ $vs.$ the dipole
fraction $s$ for a few fixed values of the dipole sizes $h$. In
accordance with physical intuition the string tension (solid lines)
decreases due to the monopole binding. Moreover, the larger the
dipole size the smaller the decrease is. This is not an unexpected
fact since the contribution of the dipole to the Wilson loop
grows with the increase of the dipole magnetic moment (equivalently,
"size" $h$ of the dipole).

Let us discuss the applicability of the pointlike dipole
approximation to a physical system, for example, to the
Georgi--Glashow model. The dipoles may be treated as
point--like particles provided the typical distance between the
plasma constituents is much larger than the dipole size $r$.  The
corresponding condition reads as follows:
\beqn
{(\rho_m + \rho_d)} \, r^3 \ll 1\,.
\label{diluteness}
\eeqn
Using eqs.(\ref{size:h},\ref{gamma}) this equation can be rewritten
as follows:
\beqn
{(1 + s)} \, h^3 \, n_m \ll 1\,,\quad n_m = \rho_m
M^{-3}_D\,,
\eeqn
where $n_m$ is the number of the monopoles in the unit Debye volume.
The string tension can be calculated in the classical approximation
(\ref{string},\ref{class2}) only in the weak coupling limit which is
equivalent to the condition $n_m \gg 1$. One can easily check that
the condition $h \ll 1$ can be fulfilled in the Georgi--Glashow model
and the correction factor (\ref{H0}) to the string tension can be
arbitrarily large. The condition $h \gg 1$ can not be fulfilled and
the result \eq{sigma:h:large} is not valid in the Georgi--Glashow
model in the weak coupling regime.

\section*{Conclusion and acknowledgments}

We have studied the confining properties of the gas of the Abelian
magnetic monopoles and the pointlike dipoles. Both the increase of
the dipole fraction at a fixed monopole density and the enlargement
of the dipole magnetic moment amplify the coefficient in front of the
linear potential ("string tension"). At the same time the monopole
binding at a fixed total number of magnetic charges leads to the
natural decrease of the string tension.

The described effects may have interesting physical applications
since the monopole--dipole gases are realized in various gauge
theories. For example, the monopole binding mechanism leads to the
high temperature deconfining phase transition in the three
dimensional compact electrodynamics and in the Georgi--Glashow
model\footnote{An alternative mechanism of the phase transition in
the Georgi--Glashow model due to magnetic vortex dynamics is
discussed recently in Ref.~\cite{Alex}.}, Ref.~\cite{AgasianZarembo}.
Another example is the electroweak model in which the formation of
the Nambu monopole--antimonopole pairs has been observed~\cite{EW} in
the low temperature (Higgs) phase. In this model the string tension
for the spatial Wilson loops is non-zero in the high temperature
(symmetric) phase in which the monopoles form a gas. At low
temperatures the string tension vanishes~\cite{EW:string}.

Author was partially supported by RFBR 99-01230a and Monbushu grants,
and CRDF award RP1-2103.

\end{document}